\documentclass[aps,prl,showpacs,twocolumn]{revtex4}
\usepackage{amssymb}
\usepackage{amsmath}
\usepackage{graphicx}

\begin{document}

\title{Spontaneous breaking of isotropy observed in the electronic transport of rare-earth tritellurides}
\date{\today}
\author{A.A.~Sinchenko$^{1,2}$, P.D.~Grigoriev$^{3}$, P.~Lejay$^{2}$ and
P.~Monceau$^{2}$}

\address{$^{1}$Kotel'nikov Institute of Radioengineering and Electronics of
RAS, Mokhovaya 11-7, 125009 Moscow, Russia}

\address{$^{2}$Univ. Grenoble Alpes, Inst. Neel, F-38042 Grenoble, France, CNRS, Inst. Neel, F-38042 Grenoble, France}

\address{$^{3}$L. D. Landau Institute for Theoretical Physics, 142432
Chernogolovka, Russia}

\begin{abstract}
We show that the isotropic conductivity in the normal state of
rare-earth tritelluride $R$Te$_3$ compounds is broken by the
occurrence of the unidirectional charge density wave (CDW) in the
($a,c$) plane below the Peierls transition temperature. In contrast
with quasi-one-dimensional systems, the resistivity anomaly
associated with the CDW transition is strong in the direction
perpendicular to the CDW wave vector $\mathbf{Q}$ ($a$-axis) and
very weak in the CDW wave vector $\mathbf{Q}$ direction ($c$-axis).
We qualitatively explain this result by calculating the electrical
conductivity for the electron dispersion with momentum-dependent CDW
gap as determined by angle-resolved photoemission spectroscopy
(ARPES).
\end{abstract}


\maketitle

\draft

Quasi two-dimensional (2D) systems with strong electronic
correlations exhibit a wide variety of properties due to strong
coupling and competition among charge, spin, orbital and lattice
degrees of freedom. One fingerprint of the resulting ground states
is often the formation of unidirectional charge or spin modulations,
as continuously investigated in manganites \cite{Sun2011}, organic
compounds \cite{Seo2000} and cuprates \cite{Kivelson2003,Vojta2009}.
There is a large debate to associate or not the high-temperature
superconductivity in the latter compounds with the existence of
charge modulation.

Very recently, a new family of quasi-2D compounds, namely rare-earth
tritellurides $R$Te$_{3}$ ($R$=Y, La, Ce, Nd, Sm, Gd, Tb, Ho, Dy,
Er, Tm) has raised an intense research activity
\cite{DiMasi95,Brouet08,Ru08}. These layered compounds have a weakly
orthorhombic crystal structure (space group $Cmcm$). They are formed
of double layers of nominally square-planar Te sheets, separated by
corrugated $R$Te slabs. In this space group, the long $b$ axis is
perpendicular to the Te planes. These systems exhibit an
incommensurate CDW through the whole $R$ series
\cite{Ru08,Lavagnini10R}, with a wave vector
$\mathbf{Q}_{CDW1}=(0,0,\sim 2/7c^{\ast})$ and a Peierls transition
temperature above 300 K for the light atoms (La, Ce, Nd). For the
heavier $R$ (Tb, Dy, Ho, Er, Tm) a second CDW occurs at low
temperature with the wave vector $\mathbf{Q}_{CDW2}=(\sim
2/7a^{\ast},0,0)$ perpendicular to $\mathbf{Q}_{CDW1}$. The
$R$Te$_3$ family can be considered as a model system in which the
structure of the CDW ground state can be theoretically studied
\cite{Yao06}. Thus a phase diagram as a function of the
electron-phonon parameter was derived with a bidirectional
(checkerboard) state if the CDW transition temperature is
sufficiently low whereas a unidirectional stripe state, as observed
experimentally, occurs when the transition temperature is higher.
This result is relevant for a deeper understanding of the charge
pattern in highly correlated materials, and particularly to the
recent determination of the biaxial CDW in underdoped cuprates
\cite{LeBoeuf2013}.

Below the Peierls transition, in all $R$Te$_3$ compounds, the Fermi
surface is partially gapped resulting in a metallic behavior at low
temperature. The layered $R$Te$_3$ compounds exhibit a large
anisotropy between the resistivity along the $b$-axis and that in
the $(a,c)$ plane, typically $\sim40$ below $T_{CDW1}$ and much
higher at low temperature \cite{Ru06}. The effect of the CDW on the
in-plane resistivity observed in experiments was very weak, no more
than few percents of the total resistance \cite{Ru08,Ru06}. However,
due to the unidirectional character of the CDW
\cite{Fang07,Lavagnini10R,Yao06}, a conductivity anisotropy in the
$(a,c)$ plane should be expected in the CDW state. In the present
paper we report the first observation and theoretical analysis of
the in-plane conductivity anisotropy in $R$Te$_{3}$ compounds below
the Peierls transition.

We have studied TbTe$_3$, DyTe$_3$ and HoTe$_3$ compounds. In
DyTe$_3$ the upper CDW appears just at room temperature at
$T_{CDW1}=302 $ K and the lower CDW at $T_{CDW2}=49$ K. In HoTe$_3$
the first and the second CDW transitions take place at
$T_{CDW1}=283$ K and $T_{CDW2}=110$ K correspondingly \cite{Ru08}.
In TbTe$_3$ the CDW ordering is observed well above room temperature
($T_{CDW1}$=336 K). The second CDW transition in this compound was
recently reported $T_{CDW2}=41$ K \cite{BB}, the lowest in the
$R$Te$_3$ series.

Single crystals of $R$Te$_{3}$ were grown by a self-flux technique
under purified argon atmosphere as described previously
\cite{SinchPRB12}. Thin single-crystal samples with a square shape
and with a thickness less than 1 $\mu $m were prepared by
micromechanical exfoliation of relatively thick crystals glued on a
sapphire substrate. The quality of selected crystals and the spatial
arrangement of crystallographic axes were controlled by X-ray
diffraction. The superlattice spots for the upper CDW were clearly
observed along the $c$-axis, demonstrating CDW ordering just in this
direction. Conductivity anisotropy measurements were performed using
the Montgomery technique \cite{Montgomery71,Logan71}.

\begin{figure}[thb]
\includegraphics[width=8cm]{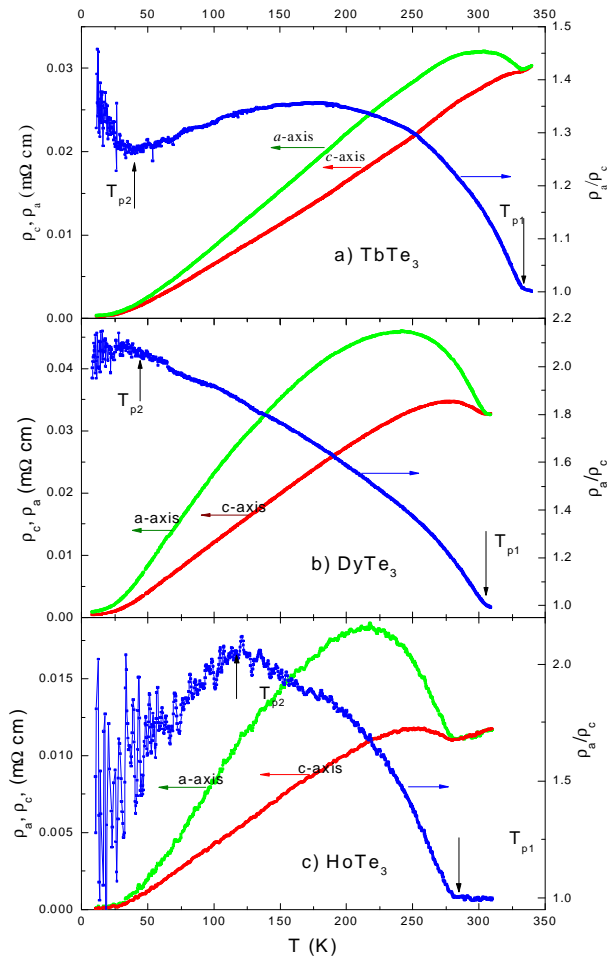}
\caption{(color online) Temperature dependence of resistivity of
a) TbTe$_3$, b) DyTe$_{3}$ and c) HoTe$_{3}$ along the $a$ and the
$c$-axis directions and conductivity anisotropy, $\rho_a/ \rho_c
$, in the $a-c$ plane.} \label{F1}
\end{figure}

Figure \ref{F1} show the temperature dependence of resistivities
of TbTe$_{3}$, DyTe$_{3}$ and HoTe$_{3}$ measured (with the help
of Montgomery method) along the $c$ and $a$ axes together with the
anisotropy ratio $\rho _{a}/\rho _{c}$ in the $ac$ plane. As can
be seen, above the Peierls transition temperature $T_{CDW1}$ all
studied compounds are practically isotropic in the $ac$ plane and
$\rho _{a}/\rho _{c}\approx 1$. Below $T_{CDW1}$ the ratio
$\rho_{a}/\rho _{c}$ strongly increases and reaches $\sim 1.4$ for
TbTe$_{3}$ and $\sim 2$ for DyTe$_{3}$ and HoTe$_{3}$ at low
temperature. The observed effect of the CDW on the resistivity
along the $c$-axis is much weaker than the change of resistivity
along the $a$-axis. Thus, the transition into the CDW state in
rare-earth tritellurides leads to a stronger increase of
resistance in the direction perpendicular to the CDW
$\mathbf{Q}$-vector than in the direction parallel to the CDW
$\mathbf{Q}$-vector. Such behavior is inverse to that observed in
quasi-one dimensional compounds with a CDW where the anisotropy is
considerable in the normal state and strongly decreases below the
Peierls transition \cite{OngBrill78}.

In DyTe$_{3}$ and HoTe$_{3}$ the observed anisotropy starts to
decrease below $T=T_{CDW2}$. On the contrary, in TbTe$_{3}$ (see
Fig. \ref{F1}) one observes a sharp increase of anisotropy below
$T=41$ K, which coincides well with $T_{CDW2}$ in this compound as
reported in Ref. \cite{BB}.

To understand the anisotropic influence of the CDW on the
resistivity along the $a$ and $c$ -axes, shown in Fig. \ref{F1}, we
calculate the electron conductivity $\sigma _{xx}=1/\rho _{a}$ and
$\sigma _{yy}=1/\rho _{c}$ in the CDW state as a function of
temperature. In the $\tau $ -approximation, the conductivity along
the main axes is given by \cite{Abrik}:
\begin{equation}
\sigma _{i}\left( T\right) =2e^{2}\tau \sum_{\boldsymbol{k}%
}\,v_{i}^{2}\left( \boldsymbol{k}\right) \left( -n_{F}^{\prime
}\left[ \varepsilon \left( \boldsymbol{k}\right) \right] \right) ,
\label{sT}
\end{equation}%
where $e$ is the electron charge, $\tau $ is the mean free time, $%
\boldsymbol{k}$ is electron momentum \cite{CommentKz}, $v_{i}$ is
the component of the electron velocity along the $i$ -direction,
$n_{F}^{\prime }(\varepsilon )=-1/\{4T\cosh ^{2}\left[ (\varepsilon
-\mu )/2T\right] \}$ is the derivative of the Fermi distribution
function, which restricts the summation over momentum to the
vicinity of FS, $\mu $ is the chemical potential, and $\varepsilon
\left( \boldsymbol{k}\right) $ is the electron dispersion. The
factor $2$ in Eq. (\ref{sT}) comes from the spin degenaracy.

\begin{figure}[thb]
\includegraphics[width=8cm]{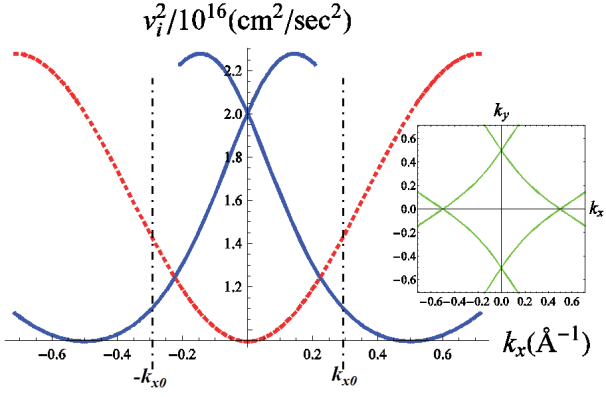}
\caption{(color online) Variation of two main components
$v_{x}^{2}$ (solid blue line) and $v_{y}^{2}$ (dashed red line) of
the electron velocity along the Fermi surface for TbTe$_{3}$ above
the CDW transition, as calculated from Eqs. (\protect\ref{Disp})
and (\protect\ref{v}) with parameters: $t_{\parallel}=2$ eV,
$t_{\perp}=0.37$ eV. Inset: Fermi surface of TbTe$_{3}$ above the
CDW transition (green solid line) calculated from Eq.
(\protect\ref{Disp}). In the CDW state the gap covers the region
$\left\vert k_{x}\right\vert \leq k_{x0}\approx
0.29\mathring{A}^{-1}$ as it shown by vertical dash-dot lines and
violates the equivalence between $v_{x}^{2}$ and $v_{y}^{2}$:
$v_{x}^{2}$ is maximum in the region under the CDW gap while
$v_{y}^{2}$ is maximum in the ungapped region.} \label{Figv2}
\end{figure}

The momentum dependence of electron velocities $v_{x}$ and $v_{y}$
can be obtained from electron dispersion with and without the CDW
gap. Without CDW the in-plane electron dispersion in $R$Te$_{3}$ is
described by a simple 2D tight binding model of the Te plane as
developed in \cite{Brouet08} in which the square net of Te atoms
forms perpendicular chains created by the in-plane $p_x$ and $p_z$
orbitals. The model parameters consist of electron hopping term
along a particular chain, $t_{\parallel}$, and perpendicular to the
chain, $t_{\perp}$. Fermi surface curvature is proportional to
$t_{\perp}/t_{\parallel}$:
\begin{equation}
\begin{split}
\varepsilon _{1}\left( k_{x},k_{y}\right) =& -2t_{\parallel }\cos
\left[
\left( k_{x}+k_{y}\right) a/2\right] \\
& -2t_{\perp }\cos \left[ \left( k_{x}-k_{y}\right) a/2\right] -E_{F}, \\
\varepsilon _{2}\left( k_{x},k_{y}\right) =& -2t_{\parallel }\cos
\left[
\left( k_{x}-k_{y}\right) a/2\right] \\
& -2t_{\perp }\cos \left[ \left( k_{x}+k_{y}\right) a/2\right]
-E_{F},
\end{split}
\label{Disp}
\end{equation}%
where the calculated parameters for TbTe$_{3}$ are $t_{\parallel
}\approx 2$ eV, $t_{\perp }\approx 0.37$ eV and $a\approx
4.4\mathring{A}$ \cite{Brouet08,Kikuchi}. These paremeters slightly
differ for other compounds of this family. The Fermi energy
$E_{F}\approx 1.48$ eV is chosen to fit the ARPES data on the FS in
TbTe$_{3}$ \cite{Brouet08,NJP}. The FS calculated above the CDW
transition from Eq. \ref {Disp} is shown in Fig. \ref{Figv2}
(inset). At $t_{\perp }=0$ the FS contains only straight lines
$k_{x}\pm k_{y}=const=\pm \left( 2/a\right) \arccos \left(
E_{F}/2t_{\parallel }\right) $, which warp at nonzero $t_{\perp }$.

\begin{figure}[thb]
\includegraphics[width=8.5cm]{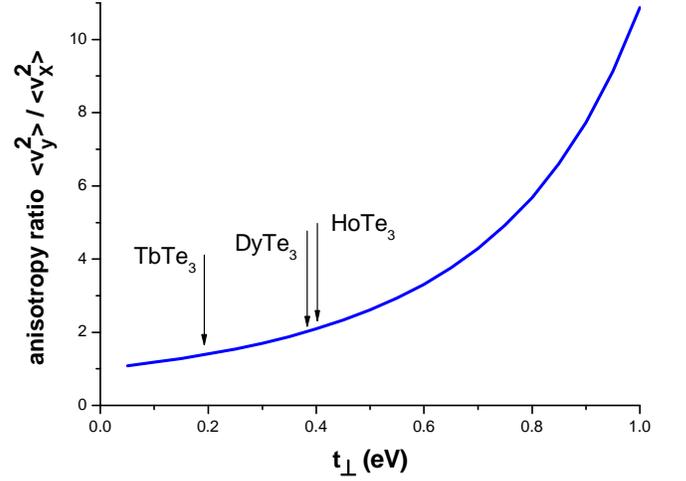}
\caption{The anisotropy ratio of the resistivity along $a$ and $c$
axis $\protect\rho_a/\protect\rho_c\approx\left\langle
v_{y}^{2}\right\rangle _{FS}/\left\langle v_{x}^{2}\right\rangle
_{FS}$ as a function of $t_{\perp }$ calculated from Eqs.
(\protect\ref{Disp})-(\protect\ref{r}) for $t_{\parallel}=2eV$.
The arrows indicate the experimental anisotropy ratio for
TbTe$_3$, DyTe$_{3}$ and HoTe$_{3}$.} \label{FigRatio}
\end{figure}

The electron velocity, calculated from Eq. \ref{Disp} using
\begin{equation}
v_{x}=\partial \varepsilon /\partial k_{x},~v_{y}=\partial
\varepsilon /\partial k_{y},  \label{v}
\end{equation}%
varies considerably along the Fermi surface. The $k_{x}$-dependence
of $v_{x}^{2}$ and $v_{y}^{2}$\ on the FS above the CDW transition
is shown in Fig. \ref{Figv2}. One can see that the maxima of
$v_{x}^{2}$ and $v_{y}^{2}$ are on different parts of the FS. This
asymmetry comes from finite $t_{\perp }$. Without CDW the summation
over momentum in Eq. (\ref{sT}) gives the same result for
$v_{x}^{2}$ and $v_{y}^{2}$, which leads to an isotropic electron
conductivity. The CDW gap violates this balance because it covers
only some parts of FS. The $k_{x}$-dependence of the CDW gap in
TbTe$_{3}$ is shown in Fig. 13 of Ref. \cite{Brouet08}. The CDW gap
is nonzero only in the region $\left\vert k_{x}/a^{\ast }\right\vert
\equiv \left\vert k_{x}a/2\pi \right\vert \leq 0.2$, which
corresponds to $\left\vert k_{x}\right\vert \leq k_{x0}\approx
0.29\mathring{A}^{-1}$. As one can see in Fig. \ref{Figv2},
$v_{x}^{2}$ has a maximum value just in this region under the CDW
gap, while $v_{y}^{2}$ is maximum in the ungapped region $\left\vert
k_{x}\right\vert >k_{x0}$. At low temperature $T\ll \Delta _{CDW}$
the integration over momentum in Eq. (\ref{sT}) includes only the
ungapped FS parts and gives the following anisotropy ratio
\cite{CommentInt}
\begin{equation}
\frac{\sigma _{yy}}{\sigma _{xx}}\approx \frac{\int_{\left\vert
k_{x}\right\vert \geq k_{x0}}dk_{x}\,\sqrt{1+\left\vert \frac{dk_{y}}{dk_{x}}%
\right\vert _{FS}^{2}}v_{y}^{2}}{\int_{\left\vert k_{x}\right\vert
\geq
k_{x0}}dk_{x}\sqrt{1+\left\vert \frac{dk_{y}}{dk_{x}}\right\vert _{FS}^{2}}%
v_{x}^{2}}  \label{r}
\end{equation}%
Taking the same parameters as for \ Eq. (\ref{Disp}) and Ref.
\cite{Brouet08} one get $\frac{\sigma _{yy}}{\sigma _{xx}}\approx
1.96$. Eq. (\ref{r}) means that the resistivity ratio $\rho
_{a}/\rho _{c}\approx \sigma _{yy}/\sigma _{xx}$ increases from $1$
to almost $2$ as the temperature decreases below $T_{CDW1}$. This
agrees well with the experimental data in Fig. \ref{F1}. The
anisotropy ratio in Eq. (\ref{r}) depends strongly on the value of
$t_{\perp }$. In Fig. \ref{FigRatio} we plot this anisotropy ratio
as function of $t_{\perp }$ with arrows corresponding to the
experimental values for TbTe$_3$, HoTe$_3$ and DyTe$_3$.

To calculate the temperature dependence of the resistivity
anisotropy one needs a detailed knowledge of the temperature
evolution of the CDW gap $\Delta \left( T,\boldsymbol{k}\right)$.
The momentum dependence of the CDW gap at the Fermi level%
\begin{equation}
\Delta \left( T,\boldsymbol{k}\right) \approx \Delta _{0}\left(
T\right) \,\Delta \left( \boldsymbol{k}\right) \approx \Delta
_{0}\left( T\right) \left( 1-k_{x}^{2}/k_{x0}^{2}\right)  \label{Dk}
\end{equation}%
is taken as a simplest fit of the experimental data in Fig. 13 of
Ref. \cite{Brouet08}. Eq. (\ref{Dk}) implies that as the temperature
decreases only the amplitude $\Delta _{0}\left( T\right) $ but not
the momentum dependence of $\Delta \left( \boldsymbol{k}\right) $
changes, which reasonably agree with ARPES data. Let take the growth
of $\Delta _{0}(T)$ at the transition temperature $T_{CDW1}$ as:
\begin{equation}
\Delta _{0}\left( T\right) \approx \Delta _{0}\left(
1-T^{2}/T_{CDW1}^{2}\right)^{\alpha}.
\label{DT1}
\end{equation}
The new electron dispersion is given by \cite{CommentED}
\begin{equation}
E\left( \boldsymbol{k}\right) =\sqrt{\varepsilon ^{2}\left( \boldsymbol{k}%
\right) +\Delta ^{2}\left( T,\boldsymbol{k}\right) }.  \label{Ed}
\end{equation}%
Since the $\boldsymbol{k}$-dependence of $\varepsilon \left( \boldsymbol{k}%
\right) $ is much stronger than that of $\Delta _{0}\left( \boldsymbol{k}%
\right) $, the electron velocity in the presence of the CDW gap is
\begin{equation}
v_{i\Delta }\left( \boldsymbol{k}\right) =\frac{\partial E\left( \boldsymbol{%
k}\right) }{\partial k_{i}}\approx \frac{\varepsilon \left( \boldsymbol{k}%
\right) ~\partial \varepsilon \left( \boldsymbol{k}\right) /\partial k_{i}}{%
\sqrt{\varepsilon ^{2}\left( \boldsymbol{k}\right) +\Delta ^{2}\left( T,%
\boldsymbol{k}\right) }}=v_{i}\left( \boldsymbol{k}\right)
\frac{\varepsilon \left( \boldsymbol{k}\right) }{E\left(
\boldsymbol{k}\right) }.  \label{vD}
\end{equation}%
Substituting this new electron velocity to Eq. (\ref{sT}), we obtain
\begin{eqnarray}
\sigma _{i}\left( T\right)  &=&\frac{e^{2}\rho _{F}\tau
}{d}\int_{-\pi /a}^{\pi /a}\frac{adk_{x}}{2\pi }\sqrt{1+\left(
\frac{dk_{y}}{dk_{x}}\right)
^{2}}v_{i}^{2}\left( k_{x}\right)   \notag \\
&&\times \int \frac{d\varepsilon }{v_{F}}\frac{-n_{F}^{\prime }\left[ \sqrt{%
\varepsilon ^{2}+\Delta ^{2}\left( T,k_{x}\right) }\right] \varepsilon ^{2}}{%
\varepsilon ^{2}+\Delta ^{2}\left( T,k_{x}\right) },  \label{sT1}
\end{eqnarray}%
where $\Delta =\Delta \left( T,k_{x}\right) $ is given by Eqs. (\ref{Dk}) - (%
\ref{DT1}), and $v_{i}\left( k_{x}\right) $ is given by Eq.
(\ref{v}) with the ungapped electron dispersion (\ref{Disp}).

\begin{figure}[thb]
\includegraphics[width=8.5cm]{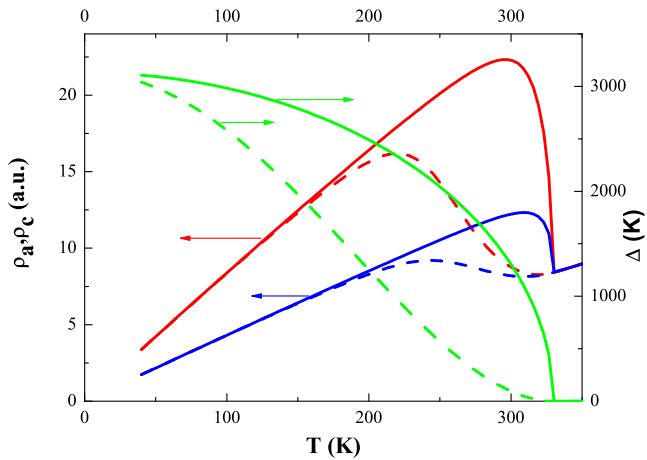}
\caption{(color online) The calculated temperature dependence of
resistivity
$\protect\rho _{a}=1/\protect\sigma _{xx}$ solid (dashed) red lines and $%
\protect\rho _{c}=1/\protect\sigma _{yy}$ solid (dashed) blue
lines in the presence of the CDW gap, $\Delta $, with the momentum
dependence given by Eq. (\protect\ref{Dk}) and with the mean-field
temperature dependence $\alpha=1/2$ in Eq. (\protect\ref{DT1})
(solid green line) and for the weaker temperature dependence near
$T_{CDW1}$ for $\protect\alpha =2$ (dashed green line). The
parameters for calculations are those for TbTe$_3$ with
$t_{\parallel}=2$ eV, $t_{\perp}=0.37$ eV, $\Delta_0=0.27$ eV,
$k_{x0}=0.29\mathring{A}^{-1}$} \label{FigRT}
\end{figure}

First, we take $\alpha=1/2$ as in the mean-field approximation.
The numerical integration of Eq. (\ref{sT1}) using
(\ref{Dk})-(\ref{vD}) gives the temperature dependence of the
in-plane resistivity along the $a$ and $c$ axis shown in Fig.
\ref{FigRT} by a solid blue line for $\rho _{c}$ and by a solid
red line for $\rho _{a}$, with the parameters corresponding to
TbTe$_3$, the same as used for Fig. \ref{Figv2}.

The comparison between experimental data (Fig. \ref{F1}a) and
calculated resistivity shown in Fig. \ref{FigRT} for TbTe$_3$
indicates that the increase of resistivity along both $a$ and $c$
axis is much more smooth than that derived from the mean field
temperature dependence of the CDW gap. Fluctuations may yield a
weaker growth of $\Delta_0(T)$ near $T_{CDW1}$ with the exponent
$\alpha$ in Eq. \ref{DT1} higher than 1/2. This dependence is
represented in Fig. \ref{FigRT} for $\alpha=2$ by a dashed green
line (to be compared with the solid green line for the mean field
prediction). The temperature dependence of the calculated
resistivity along $c$ and $a$ axis are drawn in Fig. \ref{FigRT}
with a dashed blue line for $\rho_c$ and a dashed red line for
$\rho_a$, showing a better qualitative agreement with experimental
data.

Measurements of the $T$--dependence of the CDW gap for $R$Te$_3$
compounds are scarce. Detailed measurements by ARPES on ErTe$_3$
suggest a mean-field type behavior but somewhat suppressed from the
mean-field curve \cite{Moore10}. Study of the collective modes in
DyTe$_3$ and LaTe$_3$ obtained by Raman scattering has shown that
the amplitude CDW mode develops a succession of two mean field
BCS--like transitions with different critical temperatures ascribed
to the presence of two adjacent Te planes in the crystal structure
\cite{Lavagnini10R}. The occurrence of these two transitions with
fluctuating effects between them may yield the soft growth extended
in temperature of $\Delta(T)$. Finally due to the layered structure,
the CDW gap should be anisotropic but measurements to estimate it
are still missing.

A second CDW phase transition appears of $T_{CDW2}<T_{CDW1}$. This
new transition may modify the momentum dependence of the upper CDW
and reduce the resistivity anisotropy as experimentally measured in
DyTe$_3$ (Fig. \ref{F1}b) and HoTe$_3$ (Fig. \ref{F1}c). The
increase of anisotropy below $T_{CDW2}=41$ K in TbTe$_3$ may
indicate a specific property of the low--T CDW state in this
compound, for instance a possible interference between the two CDW
distortions.

In spite of the drastic modification in the FS topology occuring at
$T_{CDW1}$ and $T_{CDW2}$ as determined by ARPES \cite{Brouet08}, a
relative small effect is induced on resistivity as seen in Fig.
\ref{F1}.  The density of states (DOS) at the Fermi level was
estimated to be suppressed in the CDW state to 77\% of the value in
the non-modificated state \cite{Brouet08,Pfuner10}; and the area of
the FS gapped by the first CDW is 3 times the area gapped by the
second one \cite{Moore10,Pfuner10}. The CDW transition at $T_{CDW2}$
is barely visible indicating a weak change in the DOS at the Fermi
level induced by the CDW transition. One possible explanation is
that, in consequence of the CDW energy gaps, the electron dispersion
(Eq. \ref{Ed} and Ref. \cite{CommentED}) gives a strong
renormalization of the quasi-particle effective mass and velocity in
the ungapped parts of the FS which would be strongly different from
that in the metallic state, as similarly described in Ref.
\cite{PDG2008}, and therefore increase the remaining DOS on the
ungapped parts of the FS.

The magnitude of the CDW gaps in $R$Te$_3$ compounds are
$\Delta_0\sim260-400$ meV leading to mean-field transition
temperatures $T_{MF}$ in the range 1500-2000 K, while the upper CDW
transition occurs between 260-400 K, leading to a large ratio
$2\Delta(0)/k_BT_{CDW}\sim10-15$ much higher than, 3.52, the BCS
value. In the conventional weak coupling Fr\"{o}hlich-Peierls model,
nesting of pieces of the FS yields the divergence of the electronic
susceptibility with the concomitant Kohn anomaly in the phonon
spectrum at the same wave vector $2k_F$. However very few CDW
systems follow this model. Similar large $2\Delta(0)/k_BT_c$ in
transition metal dichalcogenides MX$_2$ \cite{Wilson75} and in
one-dimensional systems \cite{Monceau12} were observed. For the
latter compounds strong fluctuations were believed to reduce the
critical CDW temperature below $T_{MF}$. For MX$_2$ compounds,
namely 2H-TaSe$_2$, McMillan reformulated the microscopic CDW theory
in considering that phonon over a substantial part of the Brillouin
zone (BZ) soften through the transition and that the lattice entropy
is much larger than the electronic entropy \cite{McMillan77}. While
nesting between large parts of the FS in $R$Te$_3$ compounds is
clearly seen from ARPES measurements \cite{Brouet08}, the role of
the strong electron-phonon coupling, and essentially its wave vector
dependence has been recently put into evidence to determine the
selection of the order parameter in ErTe$_3$ from Raman scattering
\cite{Eiter13}. However inelastic neutron or x-ray scattering
experiments are clearly needed for phonon spectra, especially for
the observation of phonon softening in the BZ.

In conclusion, we have shown that the quasi-isotropic conductivity
in the normal state of untwinned $R$Te$_{3}$ compounds is broken by
the CDW gap appearing below $T_{CDW1}$. The drop of conductivity is
much larger along the $a$-axis perpendicular to the CDW wave vector,
leading to a strong in-plane conductivity anisotropy (see Fig.
\ref{F1}). We explain this effect by the direct calculation of the
two components of conductivity for the electron dispersion with
momentum-dependent CDW gap as determined by ARPES . The CDW gap
covers the FS parts where the mean square electron velocity along
the a-axis is larger than along the c-axis, which leads to the
conductivity anisotropy. This conductivity anisotropy can be used
for an estimate of the electron dispersion parameter $t_{\perp }$.
We also show that the electrical anisotropy is modified when the
low-T CDW occurs below $T_{CDW2}$, which can result from interplay
between the two collective states.

We acknowledge O. Leynaud for the x-ray determination of $a$ and $c$
axis and T. F. Rosenbaum for sending us preprint of Ref. \cite{BB}
before publication. The work was supported by the Russian Foundation
for Basic Research (grants No. 11-02-01379-a and 13-02-00178-a), by
SIMTECH Program (grant no. 246937), and partially performed in the
CNRS-RAS Associated International Laboratory between CRTBT and IRE
"Physical properties of coherent electronic states in coherent
matter".


\begin{thebibliography}{99}
\bibitem{CommentKz} The dependence of $\varepsilon \left( \boldsymbol{k}%
\right) $ and $v_{i}^{2}\left( \boldsymbol{k}\right) $ on the
interlayer momentum $k_{z}$ is negligible because of strong quasi-2D
anisotropy. Hence, the summation over interplane component $k_{z}$
in Eq. (\ref{sT}) is trivial and gives $1/d$, where $d$ is the
interlayer spacing.
\bibitem{CommentInt} The integration over the momentum at $T\ll E_{F}$ is
performed in a standard way. The in-plane momentum integration is
factorized into parallel and perpendicular to FS components:
$d^{2}k=dk_{\parallel
}dk_{\perp }=dk_{\parallel }d\varepsilon /v_{F}$, where the Fermi velocity $%
v_{F}=a\sqrt{t_{\parallel }^{2}\left( 1-\cos \left[ \left( k_{x}\pm
k_{y}\right) a\right] \right) +t_{\perp }^{2}\left( 1-\cos \left[
\left( k_{x}\mp k_{y}\right) a\right] \right) }$ is independent of
$\boldsymbol{k}$
up to the second-order terms in $t_{\perp }/t_{\parallel }\ll 1$, and $%
dk_{\parallel }=\sqrt{1+\left\vert dk_{y}/dk_{x}\right\vert
_{FS}^{2}}dk_{x}$.
\bibitem{CommentED} Eq. (\ref{Ed}) is an approximation of the mean-field
dispersion in the presence of CDW with imperfect nesting, $E\left( \mathbf{k}%
\right) =\left[ \varepsilon ({\boldsymbol{k)}}+\varepsilon ({\boldsymbol{k}}-%
{\boldsymbol{Q}})\right] /2\pm \sqrt{\left[ \varepsilon ({\boldsymbol{k)}}%
-\varepsilon ({\boldsymbol{k}}-{\boldsymbol{Q}})\right]
^{2}/4+\Delta ^{2}\left( T,\boldsymbol{k}\right) }$, where $Q$ is
the CDW wave vector.
\end{thebibliography}
\end{document}